# Structural phase transitions between layered Indium Selenide for integrated photonic memory

*Tiantian Li, Yong Wang, Wei Li, Dun Mao, Chris J. Benmore, Igor Evangelista, Huadan Xing, Qiu Li, Feifan Wang, Ganesh Sivaraman, Anderson Janotti, Stephanie Law, and Tingyi Gu[*]*

Dr. T. Li, D. Mao, H. Xing, Dr. Q. Li, Dr. F. Wang, Prof. T. Gu
Department of Electrical and Computer Engineering, University of Delaware, Newark, Delaware 19716, USA
Email: tingyigu@udel.edu

Dr. Y. Wang, Dr. W. Li, Prof. A. Janotti, Prof. S. Law
Department of Materials Science and Engineering, University of Delaware, Newark, Delaware 19716, USA

Dr. W. Li
Los Alamos National Laboratory, Computer, Computational and Statistical Sciences Division, Los Alamos, New Mexico 87545, USA

Dr. C. J. Benmore
X-ray Science Division, Argonne National Laboratory, Lemont, Illinois 60439, USA

Dr. G. Sivaraman
Data Science and Learning Division, Argonne National Laboratory, Lemont, Illinois 60439, USA

The primary mechanism of optical memristive devices relies on the phase transitions between amorphous-crystalline states. The slow or energy hungry amorphous-crystalline transitions in optical phase-change materials are detrimental to the devices' scalability and performance. Leveraging the integrated photonic platform, we demonstrate a single nanosecond pulse triggered nonvolatile and reversible switching between two layered structures of indium selenide ($In_2Se_3$). High resolution pair distribution function reveals the detailed atomistic transition pathways between the layered structures. With inter-layer 'shear glide' and isosymmetric phase transition, the switching between α- and β-structural states contain low re-configurational entropy, allowing reversible switching between layered structures. Broadband refractive index contrast, optical transparency, and volumetric effect in the crystalline-crystalline phase transition are experimentally characterized in molecular beam epitaxy-grown thin films and compared to ab-initial calculations. The nonlinear resonator transmission spectra measure incremental linear loss rate of 3.3 GHz introduced by 1.5 µm long $In_2Se_3$ covered layer, resulting from the combinations of material absorption and scattering.

## 1. Introduction

Optical phase change materials (O-PCMs) have emerged as key active components in logic and memory units for optical signal processing and storage. [1] The low activation energy in chalcogenide crystalline structures offers nonvolatile tunability of material properties with versatile excitation formats. [2-6] Antimony (Sb) based chalcogenides and chalcogenide alloys have been widely adopted in all-optical and optoelectronic memory technologies, including $Ge_2Sb_2Te_5$, $Ge_2Sb_2Se_4Te_1$, $Sb_2Se_3$, and $Sb_2S_3$, [1-11] where the amorphous-to-recrystallization transitions emerge as a bottleneck for these O-PCM memories. The energy absorbed needs to heat the material beyond the melting temperature (>600°C [12-14]), followed by picosecond rapid cooling for amorphization.[15] A critical train of pulses (nanoseconds to sub-millisecond) maintains the GST temperature beyond crystallization temperature recovers the atomic ordering for data erasing. [3,16,17] During the phase transition, the competing driving forces of volume effect and the interfacial energy determine the height of the free-energy barrier for phase transitions. [18] Numerous efforts have been taken to reduce the melting temperature and crystallization time. [19-21]. In few-nm thin layers, the surface states modified atomic rearrangement reduces the crystallization timescale down to 25 ns. [22] The minimal amorphous-crystallization transition time of 0.5 ns was achieved in electronic memory, assisted by nanosecond prestructural ordering. [21] Distinguished atomic rearrangements in polymorphic layered chalcogenide materials have attracted attentions across multiple disciplines. [23-35] The low entropic crystalline-crystalline phase transitions potentially can revolutionize the power efficiency and data throughput in optical memristive devices. The electrostatically induced transitions between 2H-T' crystalline states in $MoTe_2$ are experimentally demonstrated by combining the micro-spectroscopy and electronic devices. [25] Joule heating induced reversible phase change in layered $In_2Se_3$ crystalline film has been demonstrated. [31] However, the impact of material preparation on phase change is not fully understood and the results remain controversial. [26-27, 31-35].

In this work, we investigate the atomistic transition pathways between two layered crystalline states of $In_2Se_3$ via temperature dependent high-resolution pair distribution functions (PDFs).[36]

Both α- and β-states have the same underlying rhombohedral structure system and are stable at room temperature in the form of thin film. [27,31] Their topological similarity results in low reconfiguration entropy and thus the activation energy. The activation energy, excited by photo-thermal effect or Joule heating, only needs to exceed the weak inter-layer van der Waals bonding energy and makes a relatively small rearrangement on intra-layer bonds, rather than breaking the strong covalent bonds in other O-PCM. Following the inter-layer gliding, the minor intra-layer atomic rearrangements maintain the crystal symmetry. Differential scanning calorimetry (DSC) [37] measured reversible phase transition temperatures is around 220°C, compared to the melting temperature beyond 600°C in Sb-based O-PCM. [12-14] Density functional theory (DFT) calculation shows the 0.81eV fundamental bandgap difference between the two states [38-39] promises sufficient contrast of their refractive index and conductivity, according to Moss's rule and Kramers–Kronig relations. The consistent complex refractive index spectra between DFT calculations and measurement in molecular beam epitaxial (MBE) grown continuous thin films [40-41] confirm the optical bandgaps of both states are beyond 1eV. On the device level, nonvolatile, single-shot, reversible all-optical switchings are demonstrated in MBE-grown $In_2Se_3$-silicon microring resonators (MRRs) with photo-thermal effect. The transistor arrays defined on the thin film measure thermally excited resistivity switching from $10^6$ to $10^0$ Ω cm. The polymorphic O-PCM's optical transparency at telecommunication and low entropic switching may break the power and speed bottleneck for integrated photonic in-memory computing [43-45] and reconfigurable metasurface systems. [6, 46]

**2. Low loss integrated photonic memory**

The atomic structures and optical properties of polymorphic $In_2Se_3$ were systematically explored by DFT calculations (Supporting Information Figure S1). [38] Bulk $In_2Se_3$ can exist in two main layered structures of α ($R3m$) and β ($R\bar{3}m$). Both structures consist of quintuple-layers (QL) (Se-In-Se-In-Se) linked by weak van der Waals force (Insets of Figure 1a, Supporting Information Section 1). α-$In_2Se_3$ contains both fourfold and sixfold indium atoms coordinated in tetrahedral and octahedral environments, whereas β-$In_2Se_3$ contains only sixfold indium atoms. Their similar crystalline structures result in low energy barriers and threshold excitation power for phase transitions (~220°C for thermal activation) [26-27,29,31-32,46-48]. For the α-state, the calculated optical band gap is 1.44 eV, whereas the fundamental indirect bandgap is 1.34 eV. [38] Its extinction coefficient ($k$) sharply decreases beyond 800 nm (black curve in Figure 1a). The optical bandgap for β-state is 1.27 eV, whereas the fundamental indirect band bap is significantly smaller at 0.46 eV; [38] $k$ gradually decreases from 800 nm to 1400 nm and its value is less than $10^{-5}$ at 1550 nm (red curve in Figure 1a). Both optical bandgaps are beyond 0.8 eV for transparency at telecommunication wavelengths. The contrast between fundamental or electronic bandgaps promises significant refractive index contrast towards nonvolatile optical phase-only switch and memory.

Experimentally, we explored its optical and optoelectronic responses in MBE-grown thin film and devices (method). The nonvolatile all-optical memory is demonstrated in a hybrid MBE $In_2Se_3$-silicon MRR. Here a centimeter-sized $α$-state sample is grown on a single crystalline sapphire substrate and annealed to 450ºC in the chamber [40-41] (method). The micro-Raman spectrum (black curve in Figure 1b) of the as-grown $In_2Se_3$ thin film shows a characteristic peak ~104 $cm^{-1}$ for the $α$-state. The primary peak shifts to 110 $cm^{-1}$ (red curve in Figure 1b) after thermally activated phase transition. Reverse transition to $α$-state is observed after annealing at higher temperature. The thin film can be exfoliated [49] from the sapphire substrate and transferred onto silicon MRRs for all-optical integrated photonic memory. Normal-incident laser pulses (~15 ns duration, repetition rate of 15-300k Hz, centered at 1064 nm) impinged onto the sample for triggering phase transitions in the material [50]. Reversible switching of the MRR's transmission spectra reflects the correspondent change in the refractive index of $In_2Se_3$ film (Figure 1c). The covered length of the transferred flake on the silicon single-mode waveguide is around 1.5 µm (Figure 1d). With the consideration of volumetric effect in the MBE film (Supporting Information Section 2), the effective index in such hybrid waveguides changes from 2.49 for the $α$-state, to 2.53 for the $β$-state (Figure 1e, Supporting Information Figure S2). AFM measures the flake thickness near 33.1 nm in $α$-state, and 27.9 nm in $β$-state (Supporting information S3). The reduced thickness in $β$-state partially offsets its positive contribution to the effective index. The intermediate resonance wavelength shifts and extinction ratio changes illustrate the refractive index variations with single exposure pulse peak intensity (repetition rate of 20 kHz, Figure 1f). Transition into $β$-state was achieved at the exposure energy of 0.25 nJ. The resonance wavelength red-shifted 100 pm. The extinction ratio increased from 4.45dB to 6.27dB. With the linear coupled-mode-theory fitting of those transmission spectra, the extracted intrinsic quality factors were switched between 4,800 ($α$-state) and 7,500 ($β$-state), and the coupling quality factor was kept around 5,000. The linear loss reduction may indicate suppressed scattering loss in the mechanically reformed $β$-state MBE flake. [51] When the exposure energy increased to around 0.56 nJ, the resonance peak shifted back, and the extinction ratio decreased to the original value. Higher single pulse energy at 0.7 nJ led to melting and amorphization, the resonance spectrum blue-shifted to the smaller wavelength compared to the initial $α$-state (Figure 1f). The micro-Raman verified the $In_2Se_3$ flake's state after each laser exposure. The three-dimensional temperature distribution evolvement under laser excitation is detailed in

Figure S4a-d. The higher transition temperature for *β*- to *α*-state leads to nearly doubled optical switching energy compared to *α*- to *β*- state transition. The repeatability of the optical switcher has also been demonstrated in the other device at higher repetition rate of 300 kHz (Figure 1g). 30 pulses exposure results in the similar resonance shift and extinction ratio change (supporting Information Figure S4e). The retention and cyclability of the device can be improved with a cladding layer preventing oxidation of the chalcogenide material. The additional loss in the hybrid device is composed of scattering and defects absorption. As only the linear defects absorption contributes to photothermal dispersion, we utilize the ultrasensitive integrated photonic nonlinear spectroscopies to extract each contribution (Supporting Information Figure S5 and Table S2). The linear material absorption rate increases from 16.7 GHz to 20 GHz for the 1.5 µm coverage, and the total scattering loss rate keeps around 20 GHz for both monolithic and hybrid devices. The ratio between scattering and linear absorption depends on the van der Waal contact between the crystalline $In_2Se_3$ thin film and silicon photonic substrate, and thus vary among devices (Supporting Information Table S3).

**3. Phase transitions between layered structures**

DSC, in-situ high-energy XRD and DFT combination may quantitatively illustrate the atomic structure evolvement during the phase transitions. With initial and final states confirmed by Micro-Raman spectroscopies, DSC was used to analyze the thermal storage dynamics in the intermediate temperatures (Supporting Information S4). 25 mg high purity *α*-state $In_2Se_3$ power was enclosed in a container, with temperature cycled between 25°C to 350°C for multiple loops at the rate of 20°C/min. The repeating endothermic (220°C) and exothermic (190°C) peaks indicate the phase transitions during heating and cooling (Figure 2a). The small temperature difference between the onset and peak temperatures in the DSC measurement (Figure. 2a) leads to the discontinuity in the first derivative of the free energy, and thus the observed structural transitions are the first-order phase transitions. [52] In-situ high-energy XRD with a source wavelength of 0.124014 Å captures the subtle bond length transition across the layered structures (methods and Supporting Information Figure S8a-c). [53] The sample was mounted in a Linkam furnace for temperature control. Simultaneously, the XRD spectra were recorded at each temperature incremental of 2°C. Figure 2b shows the resulting structure factors *F(Q)*. Near the phase transition temperature of 220°C ($T_c$), the peak intensities of converted reduced structure factors at $Q$ = 2.01 Å$^{-1}$ and 2.13 Å$^{-1}$ exhibit clear transitions (Supporting Information Figure S8c). The corresponding PDF data at selected temperatures with 40°C intervals revealed variations in inter-atomic distances and atomic disorder (Figure 2c). Mid- and long-range orders confirm crystalline states at all temperatures (Supporting Information Figure S8d).

The atomistic pictures describing the detailed transition process between the layered structures (insets in Figure 1a) are illustrated in Figure 2d-e. In the $α$-state, the outer Se-atoms between QLs are aligned, whereas in the $β$-state they are located at the interstitial sites of the Se-atoms in the neighboring layers. The structural transition can be initiated by 'interlayer shear glide', where each QL layer is structurally the same, but the layers are displaced with respect to each other (Figure 2d). [54] The QL-QL shear gliding is facilitated when the thermal activation energy exceeds the low van der Waals bonds energy, resulting inter-QL distance compression as the outer Se falls into the interstitial sites (Figure 2d).

Short-range PDFs quantify the inter-atom distances change after the intra-QL rearrangement (from Figure 2d-e). The detailed calculations of inter atom pair distance distributions provides break downs of contributions to each PDF peak (Figure S9a). Side-views of the atomic structures illustrate the intra-QL bond twists from the Wurtzite-type in $α$-state QL (right inset in Figure 2c) to the face-centred cubic (fcc) in $β$-state $In_2Se_3$ (middle inset in Figure 2c). The noticeable transition between similar topologies involves shortened In-In distance (In-$In_2$). The trends of those distance shifts align with the measured short distance PDFs (Figure S9b). At room temperature, the PDF peak at the interatomic distance ($r$) of 2.64Å correspond to In-Se, and the one at $r$ = 4.03 Å is attributed to In-In and Se-Se bonds (Figure 2c). [35] Abrupt PDF transition takes place the temperature increased from 210°C (grey curve) to 250°C (light pink curve in Figure 2c). Compressions of 0.38% and 1.0% are derived for the In-Se (Figure 2f) and In-In (and overlapping Se-Se) distances (Figure 2g), respectively. The lattice constant difference is consistent with the calculated [30, 38] and measured [55-56] values between $α$- and $β$-states. The PDF peak near $r$ = 4.80 Å has contributions from both In-Se and In-In pairs (In-$In_2$) at $α$-state (black curves), but its intensity decreases after transition to $β$-state with only In-Se pair contribution (red curve). The In-In pair distance near $r$ = 4.80 Å (In-$In_2$) is shortened, and merged with the PDF peak on the left (for In-$In_1$), which results in expanded FWHM (6.4%) for the peak near $r$ = 4.03 Å (Figure 2g). The phase transition leads to 7.5% disorder reduction in In-Se bond (thermal offsets excluded). The trend of In-Se bond disorder is similar to In-In correlations for the cooling process.

## 4. Complex refractive index spectra and volumetric effects

The complex refractive index spectra were characterized in centimeter-sized MBE grown samples and compared to theoretical values (Supporting Information Figure S10). Crystalline $α$-phase $In_2Se_3$ grows on sapphire with BiInSe wetting layers (method). Multi-spot micro-Raman spectra indicate that the as-grown thin film contains a small fraction of $β$-phase, so the measured $n$ value (solid blue curve in Figure 3a) is slightly higher than the calculated value (dashed blue curve in Figure 3a). The extinction coefficient dramatically reduces below the optical bandgap of 1.44 eV. The band-to-band transition is indicated in the calculated band diagrams (inset of Figure 3a). A fully converted β-phase can be obtained by treating the sample with rapid thermal annealing (RTA) (method). The measured $n$ aligns with the calculated value (blue curves in Figure 3b). The measured $k$ shows the same trend as the calculations (red curves in Figure 3b). The broadband offsets (~0.07) between the measurement (dashed) and calculated value (solid red curve in Figure 3b) might be attributed to the stress-induced surface roughness and scattering loss. This surface roughness is caused by different thermal expansion coefficients between the substrate and the epitaxially grown thin film. The stress-induced ripples should be reduced in $In_2Se_3$ flakes, as reduced linear loss is observed in the phase changed hybrid device (Figure 2a). [49] It is noted that the measured $k$ at 1064 nm for α- ($k=0.011$) and β- states ($k=0.032$ after excluding scattering loss) are beyond the theoretical values (Figure 1a), due to bismuth (Bi) doping (Discussion section). [41] Figure 3c plots the temperature-dependent refractive index at 1550 nm versus correspondent thin film thickness with temperature-controlled states. XRD verifies that the as-grown sample is a mixture state (black dots in Figure 3d), as the 18.35º and 18.80º are identified as α- [004] and β- [006] peak [57] (Supporting Information Figure S11). Lorentzian fitting obtains the same line width of 0.14º for both peaks, indicating high-quality crystalline states (black solid curves in Figure 3d). After RTA treatment, the α-peak disappears (red dots in Figure 3d). The β-peak slightly shifts to 18.86º, with an expanded linewidth of 0.204º. The optical image (inset of Figure 3d) captures the expansion of the phase changed area. Those β-phase domains expand with RTA time and temperature. Eventually, the boundaries of the phase changed domains merged and became a continuous sheet, with the stress-induced surface roughness of 15 nm.

## 5. Resistivity switching by thermal, electric and optical field excitations

To better understand the electronic response in MBE $In_2Se_3$ during phase transition, temperature-dependent in-situ resistivity testing is firstly carried out in arrays of micro-transistors (method). The sample is mounted on an in-house-made heater with temperature cycled between 25ºC and 500ºC, with its current-voltage curve at low voltage (V= 0.8V) simultaneously monitored. The derived resistivity drops steeply near 350ºC. Near the six orders of magnitude resistivity reduction remained unchanged during the cooling process (Figure 4a). At room temperature, the resistivity of β-$In_2Se_3$ (~40 Ω cm) with a fundamental bandgap of 0.46eV (Supporting Information Figure S1e) is comparable to Germanium (47 Ω cm) with a fundamental bandgap of 0.67eV. The contact resistance is excluded from those results (Supporting Information Figure S13). At room temperature, we observed resistivity switching (from $10^6$ Ω cm to $10^3$ Ω cm) and hysteresis loop (inset in Figure 4a). Low-voltage characterizations confirmed three orders of magnitude resistivity switching by the Joule heating (Supporting Information Figure S14). The threshold voltage and contact resistance can be significantly improved with alternative metal contact (indium or graphene) [58, 31] or vertical junction with large contact area. [31]

Similar resistivity switching is also observed under ultrafast pulsed laser excitations (black dots in Figure 4b). The excitation light centered at 1550 nm with 90 femtosecond duration was guided by a single-mode fiber and launched onto the active area between the electrode from the top (Inset of Figure 1b). The resistivity reduced three orders of magnitude as the peak intensity of the pulsed excitation reaches 1 GW/cm$^2$ (average power of 10 kW/cm$^2$). At the same average power, only small resistivity fluctuation is observed under continuous wave (CW) excitation (grey empty squares in Figure 4b). Femtosecond pulse excitation generates free carriers via two-photon absorption, and the related thermal effect is weaker compared to CW excitation. The absorption saturation effect from the free carriers leads to the strong power dependence of the photoresponse. The pulsed light induced resistance change exhibits weak bias dependence in the range from 2 to 16V. The photoresponse reduces as the lateral bias voltage is less than 2V (Supporting Information Figure S15). The bias dependence might indicate the contributions from ferroelectric response [40, 42, 59, 60] or interfacial charge effect [61]. Additional experiments, especially gate-bias dependent photocurrent, are required to differentiate their contributions to the overall photocurrent.

## 6. Statistical Analysis

We demonstrate the nonvolatile phase change performance of In$_2$Se$_3$ from thermal, electrical and optical aspects. Micro-Raman and XRD verify the films' crystalline state. Multiple measurements indicate the micro-Raman peak wavenumber variation is within ±0.5 cm$^{-1}$. Each data point in XRD measurement (Figure 2c&3d) are captured with sufficient accumulation time. The random noise is cancelled out and the data overlap between measurements. The repeatability of DSC measurement is also verified.

The transmission spectra of In$_2$Se$_3$/silicon microring devices in Figure 1c are normalized with the ones for silicon waveguides with grating couplers. The loss and the wavelength dependency of the grating coupler is subtracted from the transmission spectrum. Same normalization method is applied to the on-resonance transmission in Figure 1g. The percentage of the optical energy exposed on the active area (Figure 1f) is calculated based on the estimated spot size of 10 µm, with standard deviation of ±10%. For extracting the resonance wavelength, extinction ratio, absorption loss and the scattering loss of the In$_2$Se$_3$/silicon device (Figure 1f, Figure S5), the transmission spectra are fitted by the coupling-mode-theory programmed in MATLAB. Four hybrid samples are fabricated to exam the repeatability (Table S3).

The refractive index data in Figure 3a&b is collected by J.A. Woolam M-2000VI, and modeled using the software of CompleteEASE. The current versus voltage data in Figure 4 are all collected with source meter under high precision mode, then the corresponding resistivity is calculated according to the IV curve by MATLAB. The optical power is fed into the device with single mode fiber. The average illumination power is estimated in the same way as the method described for Figure 1f.

## 7. Discussion

It is noted that the phase transition pathways heavily depend on atomic disorders in most material systems, including grain boundaries, dopants, and interface states. [7] Controversial phase transition mechanisms and processes have been reported in $In_2Se_3$ samples (Supporting Information Table S4), which might be attributed to the material growth of the processing procedures. Here we observed higher phase transition temperatures in the MBE sample (400°C-600°C) compared to the pristine power form (~220°C). The discrepancy might be attributed to low density Bi doping in the MBE sample. [62] The unintentional Bi doping is likely to be diffused from seed layer during the epitaxial growth. Suppression of the Bi diffusion may reduce the phase transition temperature to the value comparable to the high purity powder form (Figure 2a). Given the limited literature, a systematic study of the correlations between the impurities and the phase change temperatures can guide the experimental pathways of material preparation and the device geometry design. On the other hand, the Bi diffusion reduces the bandgap of the material, allowing higher absorption coefficient across the visible and near infrared range.

The phase change mechanism between the layered $In_2Se_3$ is distinguished from the amorphous-crystalline transitions from the following two aspects: (1) the threshold activation energy only needs to exceed the weak inter-QL van der Waals bond energy to trigger the inter-QL shear-gliding for phase transition; (2) The subsequent intra-QL atomic twist (Figure 2c insets) does not involve any major bond rearrangement, such as rotation or reconnection; (3) The inter-layered structure transitions do not breakdown the crystal symmetry. The low re-configurational entropy between layered structures promises energy efficiency and cyclability advancement of the layered $In_2Se_3$ for nonvolatile photonic switch. The reverse $\beta$- to $\alpha$-state transition is likely to be initiated by the similar mechanism: broken van der Waal bond and shear-glide, followed by the intra-QL arrangement (reverse transition from fcc to Wurtzite-type). The film thickness expansion in the reverse $\beta$- to $\alpha$-state transition shows the reverse volumetric effect, as the interstitially aligned QLs shifts to the aligned QLs with 10% incremental inter-QL distance. [54]

The DFT calculations show the volumetric contrast of 8%, compared to the AFM measured flake thickness change of 15±6% (Supporting Information Figure S3).

## 8. Conclusion

In conclusion, we explored the polymorphic layered material of In$_2$Se$_3$ for optical memory. The low entropic phase transitions between layered structures are potential for manifesting the energy efficiency and speed limitations of integrated photonic memoristive devices and circuits (Supporting Information Table S5), with applications from optical buffer to in-memory computing. In contrast to the microsecond to millisecond amorphous-crystallization transitions, we experimentally demonstrate that only a single nanosecond pulse with nearly double energy recovers the material from $\beta$- to $\alpha$-state. The optical transparency for both states of In$_2$Se$_3$ at telecommunication wavelengths allows low insertion loss and phase-only memory device. The measurement in thin-film MBE samples verified the calculated complex refractive index spectra. The structural transitions between layered states are investigated by combinations of high-energy XRD and DSC measurement. The optical and optoelectronic properties in thin-film MBE grown samples are carefully evaluated at the device level, including the demonstration of all-optical nonvolatile switching in hybrid MRR, thermal excited resistivity switching up to $10^6$, and hysteresis loop in microscale transistors. Especially, the refractive index of silicon is between those two crystalline states of In$_2$Se$_3$, which enables versatile designs for silicon photonic modulators and integrated transformative optics. [46]

## 9. Methods

*Computational Methods.* The calculations are based on density functional theory [63-64] and the screened hybrid functional of Heyd, Scuseria, and Ernzerhof (HSE06) [65] as implemented in the Vienna ab initio simulation package (VASP) [66]. The interactions of valence electrons with the ionic cores are treated using the projector augmented wave (PAW) potentials. [67] The reciprocal space integration was performed using a mesh of Γ centered 6x6x6 special *k* points for structural optimization and a mesh of 8x8x8 for the calculation of the frequency-dependent dielectric functions.

*Characterization of crystalline-crystalline phase transition.* XRD data were collected at beamline 6-ID-D at the Advanced Photon Source, Argonne National Laboratory on a WAXS detector (Varex CT4343) using 99.9758 keV ($\lambda$ = 0.124014 Å) x rays. The detector has an active area of 2880×2880 pixels, with a pixel size of 150 µm. The sample to detector distance is ~362 mm. High purity samples (Aldrich, 99.99% trace metal purity) were mounted on Linkam furnace and heated from room temperature to ~350 °C. Calibration of the detector distance, beam center, detector tilt, and rotation was performed using the FIT2D and GSAS-II software package based on the measurement of a CeO$_2$ NIST standard.[68] After detector calibration and 2D images integration, reduction of the 2D images to 1D diffraction patterns yielded the x-ray intensities. PDF function was computed from the 1D diffraction patterns at different temperatures. [62] DSC measurement was carried out in TA instrument Discovery Q600. Weighted In$_2$Se$_3$ powder samples were conducted thermograms under a standard nitrogen flow of 100 mL/min. The heat flow rate in a reference sample was simultaneously monitored at the same ramping rate for improving accuracy.

*MBE material preparation and characterization.* Here 50 nm In$_2$Se$_3$ films are grown on c-plane sapphire in a dedicated Veeco GenXplor MBE chamber following the procedure outlined in the reference. [69] The thin film MBE samples were characterized by a micro-Raman spectrometer, XRD (Bruker D8 with coupled θ-2θ scan for picking up planes in the c-direction), and ellipsometer (J. A. Woolam M-2000VI with a wavelength range of 370 – 1690 nm and spectral resolution of 2.2 nm). RTA is carried out in a nitrogen environment to minimize oxidation.

*Silicon photonic device fabrication.* The MRRs were fabricated on an SOI (111) substrate from Soitec, with 220 nm device layer on 3 μm silicon dioxide layer. The designed patterns were defined in CSAR 6200.09 positive resist using a Vistec EBPG5200 electron beam lithography system with 100kV acceleration voltage, followed by optimized resist development and single-step dry etch procedures.

*MBE sample transferring.* Thermal release tape is first applied onto the as-grown MBE chalcogenide film on sapphire substrates. The film can be easily peeled off with the tape. Microring devices were patterned on SOI substrate. Then a thin layer of PMMA was spin-coated on the SOI substrate and patterned with e-beam lithography for exposing the MRR and preventing residual adhesion on other areas during transferring. The film adhered to tape is then applied onto the patterned SOI substrate. Bubbles might be found on the transferred interface, which can be removed by applying pressure to push the bubble onto the edge. A cotton bud is recommended for the bubble removal, as ridged tools, such as tweezers, can break the crystalline structures. The sample is baked at 110ºC for tens of seconds. The thermal release tape is cured and detached from the MBE film, leaving the sample onto the target substrate. A final acetone bath removes PMMA and unwanted flakes.

*Electronic device fabrication and testing.* Wedge-shape electrodes were defined on In$_2$Se$_3$/sapphire in a poly methyl methacrylate resist using electron-beam lithography. The active regions are 10μm-wide and 1.5 μm long (Inset of Figure 4b). After development, chrome/gold metal layers (10 nm/90 nm) were deposited using electron-beam evaporation. 4155B Agilent semiconductor parameter analyzer measures voltage-dependent current. A source meter (Keithley 2000) records photocurrent under bias. The source for pulsed photocurrent measurement is a femtosecond laser centered at 1550 nm with a duration less than 90 fs and a repetition rate of 55MHz (Calmar laser CFL-10CFF).

*Near infrared laser induced phase transitions.* A Spectra-Physics HIPPO Nd:YAG laser was mainly used, which can generate laser pulses with 1064 nm wavelength and ~15 ns pulse duration. Bi doping in the material allows effective absorption at 1064nm for photothermal effect and phase transition. The laser device was connected to a computer, where the pulse repetition rate (15 kHz-300 kHz) and diode current can be set on a control panel. A low power (maximum power < 4 mW) HeNe laser (633 nm) was used for optical alignment as a visible laser and a collimated white light can illuminate the sample so that the work region can be seen on the screen. All three sources were focused on the sample using a long working distance NIR objective (NA=0.42). An optical attenuator was placed in front of the pulse laser to adjust the laser power.

**Supporting Information**

Supporting Information is available from the Wiley Online Library.

**Conflict of Interest**

The authors declare no conflict of interest.

**Acknowledgments**

This work is funded by Army Research Office (W911NF2010078YIP). We thank M. Gerhold for the discussions of the project. Y.W. and S.L. acknowledge support from the U.S. Department of Energy, Office of Science, Office of Basic Energy Sciences (DE-SC0016380). W. L. and A. J. are supported by NSF Early Career Development Program DMR-1652994, and made use of the XSEDE supercomputer facilities, National Science Foundation Grant No. ACI-1053575. W. L. is partially supported by the Laboratory Directed Research and Development Program of Los Alamos National Laboratory (LANL) under project number 20210087DR. LANL is operated by Triad National Security, LLC, for the National Nuclear Security Administration of the U.S. Department of Energy (Contract No. 89233218CNA000001).

**Figures**

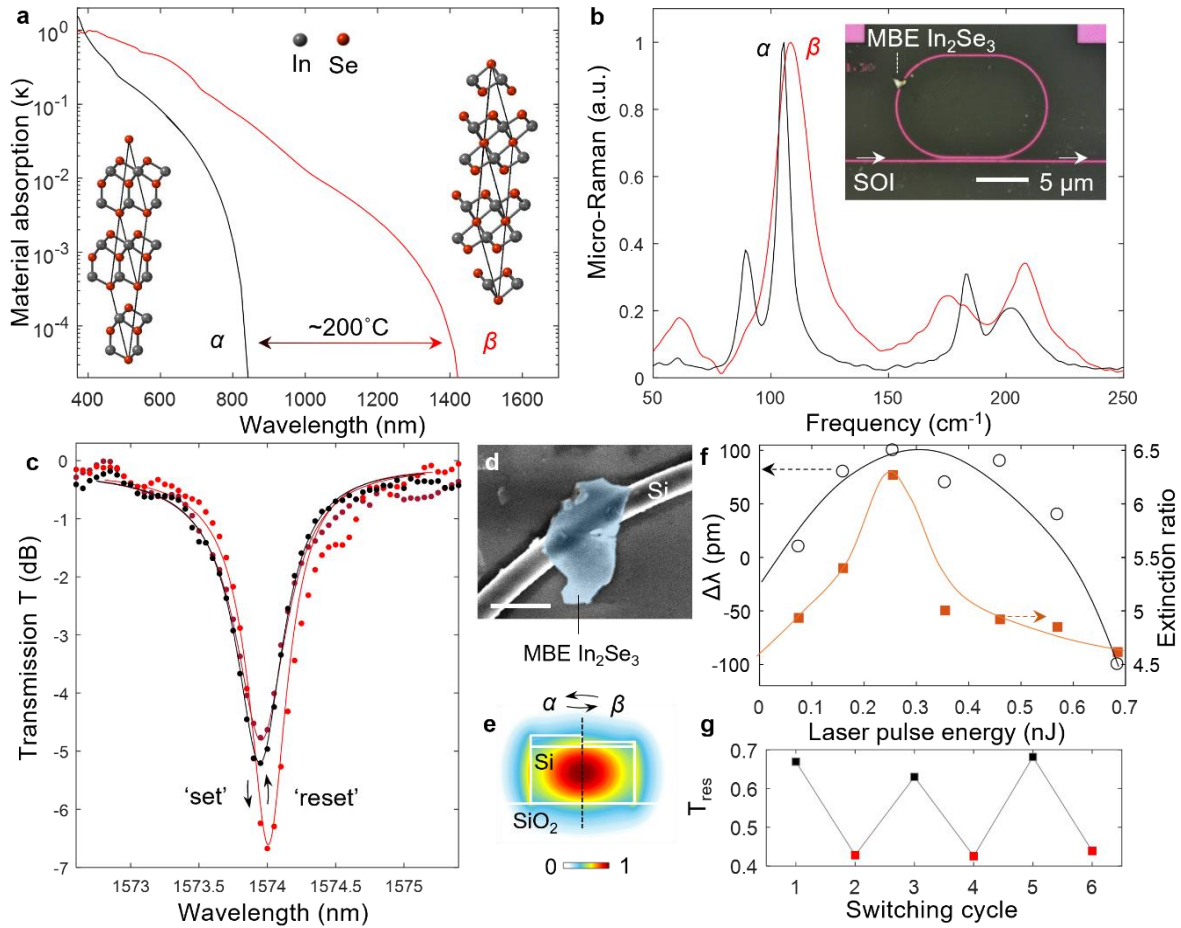

**Figure 1.** Nonvolatile all-optical memory in epitaxial In$_2$Se$_3$-silicon microring resonators. a, Calculated extinction coefficient spectra for the two layered crystalline states in In$_2$Se$_3$. Insets: Correspondent at Rhombohedral primitive cells for $\alpha$- and $\beta$-states. b, Micro-Raman spectra of MBE grown thin film In$_2$Se$_3$. Inset: the optical microscope image for In$_2$Se$_3$-Si microring device. c, Normalized transmission spectra for the hybrid resonator with $\alpha$- (black), $\beta$- (red), and retrieved $\alpha$-state (dark red) In$_2$Se$_3$. The dots are experimental data, and the curves are coupled-mode-theory fittings. d, As-prepared MBE film transferred on silicon photonic single mode waveguide. AFM measures the film thickness to be ~50 nm. Scale bar: 1 µm. e, The mode profiles for $\alpha$-In$_2$Se$_3$ (left) and $\beta$-In$_2$Se$_3$ (right) on Si WG are placed side-by-side to illustrate the volumetric effect in PCM. f, Measured resonance shift (black empty circles) and extinction ratio (orange filled squares) versus estimated single pulse energy (controlled via intensity with fixed duration). The curves are eye guiders. g, On-resonance transmission for three switching cycles with multiple pulse exposure.

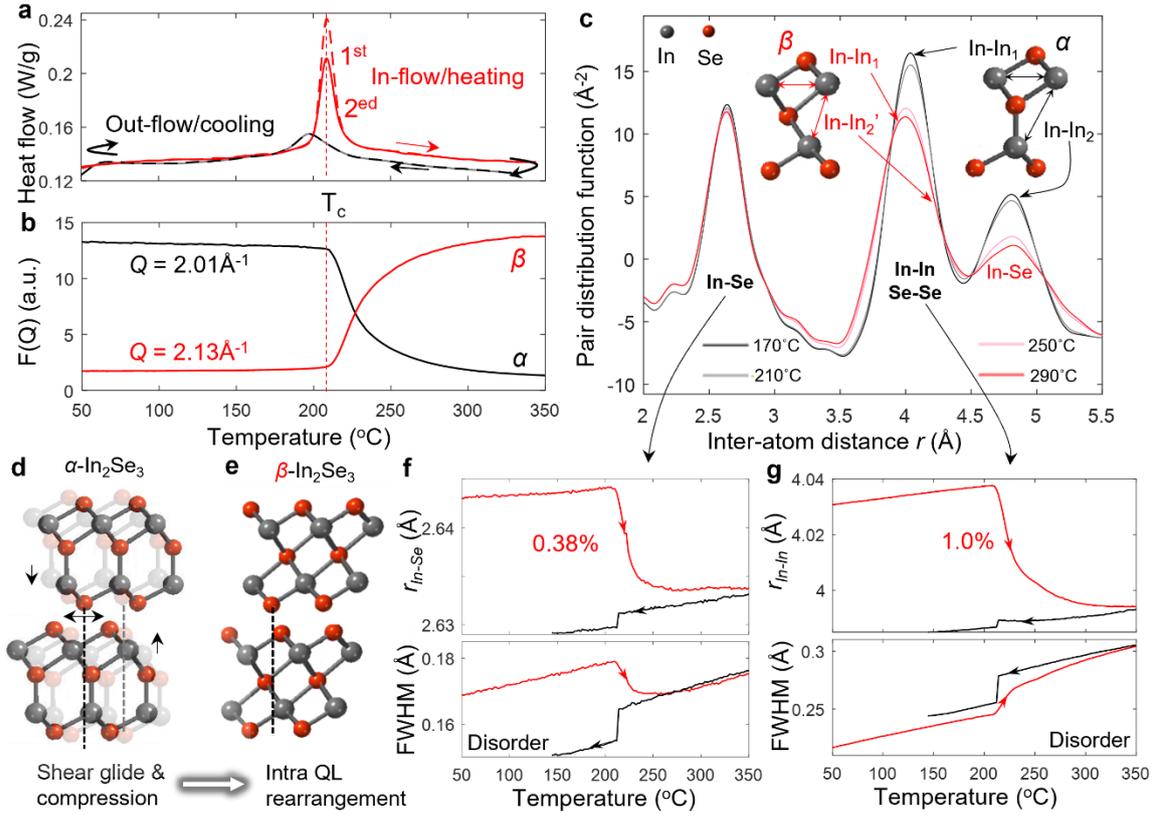

**Figure 2.** Structural transition process in layered Indium (III) Selenide materials. a, Measured heat flow dynamics by differential scanning calorimetry. The red and black curves show endothermic (220°C) and exothermic (190°C) peaks during heating and cooling, respectively. The sample temperature is cycled between 25°C to 350°C for two loops, at the rate of 20°C/min. b, Correspondent high-energy X-ray diffraction, named reduced structure factors $F(Q)$, at the scattering momentum $Q$ of 2.01Å$^{-1}$ (black) and 2.13 Å$^{-1}$ (red). c, The temperature dependent pair distribution function with 40°C per step during heating. Middle/ right insets: side views of one QLs in $α/β$ state. d, Atomic structure of $α$-In$_2$Se$_3$ at room temperature (semi-transparent) and after thermally activated shear-glide (solid color). The outer Se atoms fall into the interstitial sites, which results in compression of inter-layer distance. e, Subsequent intra-QL rearrangement (Insets in c) results in the structures as $β$-In$_2$Se$_3$. Only the middle two QLs are included here for clarity. f, The extracted inter-atom distance and FWHM from the PDF spectra in c at ramping (red curves) and falling (black curve) temperature for In-Se and g, In-In/Se-Se pairs.

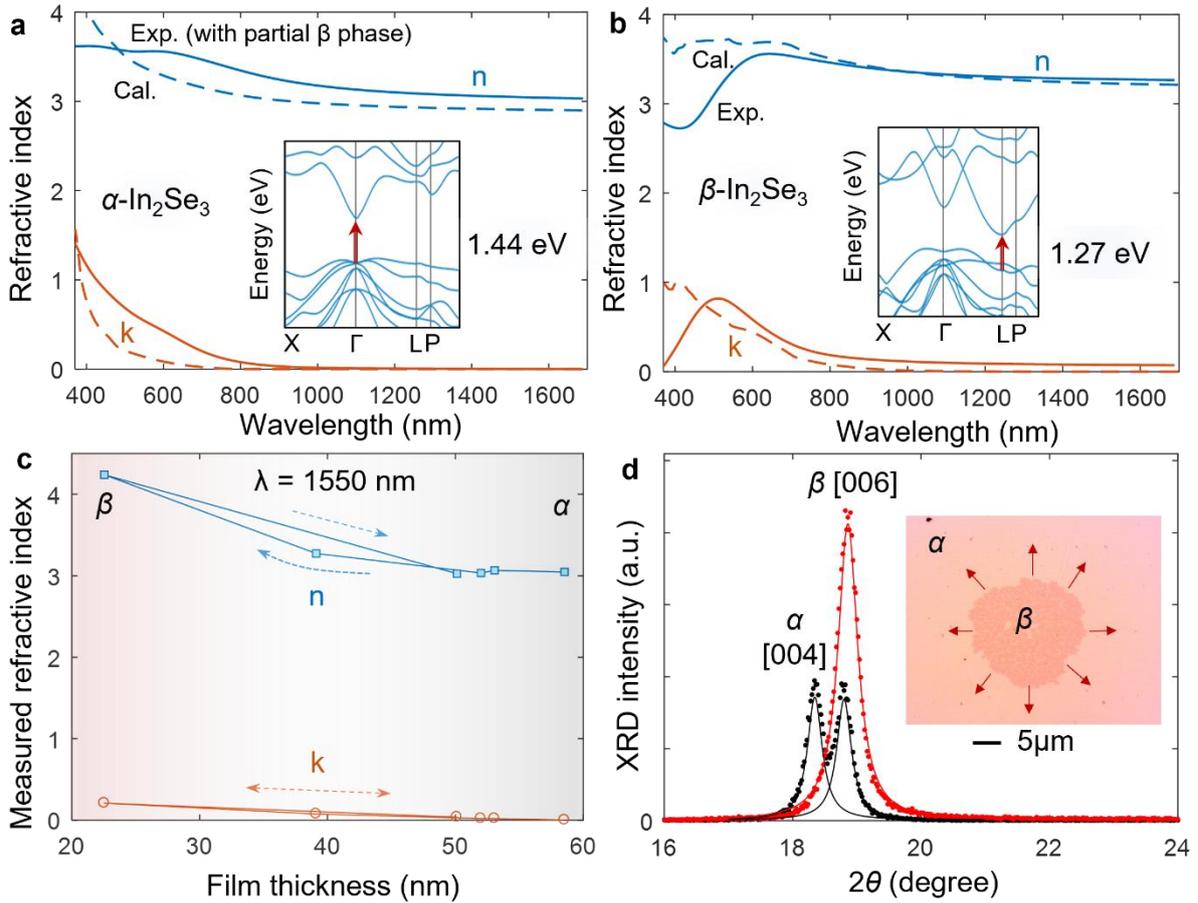

**Figure 3.** Complex refractive indexes in MBE grown *α*- and *β*-In$_2$Se$_3$, with thermally activated phase transitions. a, Comparison of measured (solid curves) and DFT calculated (dashed curves) refractive index spectra in *α*- and b *β*-In$_2$Se$_3$. The blue and red curves represent real and imaginary part of the complex refractive index. Inset in a: Band diagrams of the *α*-In$_2$Se$_3$ with optical bandgap of 1.44 eV, and b: *β*-In$_2$Se$_3$ with optical bandgap of 1.27 eV. c, Measured reversible refractive index switching versus film thickness at 1550 nm. d, XRD measurement reveal the crystalline structure transitions in the film. Inset: Top view of the emerging *β*-phase with expanding boundaries, captured by an optical microscope image.

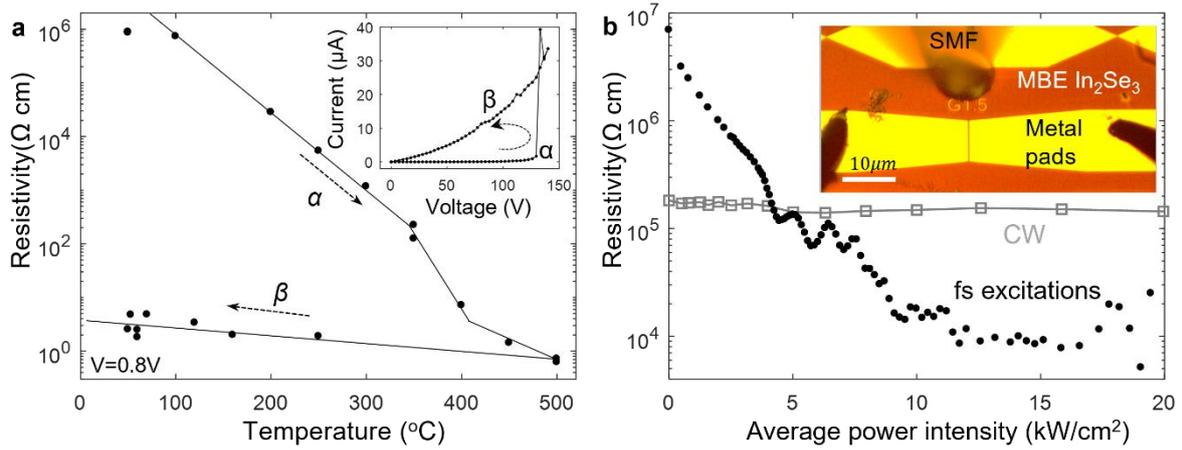

**Figure 4.** Resistance switching in MBE In$_2$Se$_3$ thin film with thermal, electric field, and pulsed laser excitations. a, In-situ measurement of resistance with ramping and declining substrate temperature. The lines are eye guiders. Inset: Hysteresis loop of a transistor defined on the thin film. b, Resistance versus excitation optical power centered at 1550 nm. The bias voltage is set at 5V for both measurements. Inset: optical microscope image of the device under test. A single mode fiber (SMF) guides the light from tunable laser onto the active area between the metal pads.

Supporting Information

# Structural phase transitions between layered Indium Selenide for integrated photonic memory

*Tiantian Li, Yong Wang, Wei Li, Dun Mao, Chris J. Benmore, Igor Evangelista, Huadan Xing, Qiu Li, Feifan Wang, Ganesh Sivaraman, Anderson Janotti, Stephanie Law, and Tingyi Gu[*]*

## S1. Calculation complex refractive index spectra

$In_2Se_3$ can exist in multiple crystalline states, among which $α$- and $β$- are layered structures and the $γ$-phase has a three-dimensional structure. Numerous works have been published for studying different crystalline formats in $In_2Se_3$. Four crystalline phases have been reported, and the calculated electronic band structures and measured bond lengths display important differences [S1-S4]. The layered structure of $α$- and $β$-state $In_2Se_3$ are strongly bonded in the a-b plane, with weak van der Waals bonding between unit cells in the c direction (Figure S1a-c). Here we use the screened hybrid functional of Heyd-Scuseria-Ernzerhof (HSE06) hybrid functional for calculating the band structures. Figure S1d-e shows the calculated band diagram of $α$- and $β$-states between the energy level from -2 to 4 eV, with reference energy level set at the valance-band maximum.

The complex refractive index spectra (Figure S1g-i) can be derived from the band structures given in Figure S1d-f. We first computed the complex frequency-dependent tensor, and then the complex refractive index spectra were derived from the optical transition matrix elements between valence and conduction band states.

$$n = \sqrt{\frac{|\varepsilon_r + i\varepsilon_i| + \varepsilon_r}{2}} \quad \text{(S-1a)}$$

$$k = \sqrt{\frac{|\varepsilon_r + i\varepsilon_i| - \varepsilon_r}{2}} \quad \text{(S-1b)}$$

The real (*n*) and imaginary (*k*) parts of the refractive index in the range from 400 nm to 1700 nm are derived from the band diagram for $α$-, $β$- and $γ$-states (Figure S1g-i). For $α$- and $β$-$In_2Se_3$, the anisotropic in-plane ( $//$ ) and out-of-plane ( $\perp$ ) refractive index spectra are origniated from the anisotropy electronic screening in those layered structures (Figure S1g-h). $γ$-state is isotropic (Figure S1i). The calculated optical bandgaps for $α$-, $β$- and $γ$-states are 1.44eV, 1.27eV, and 1.8 eV, respectively [38]. These energies are all beyond the photon energy in telecommunication wavelength (~0.8eV) [38].

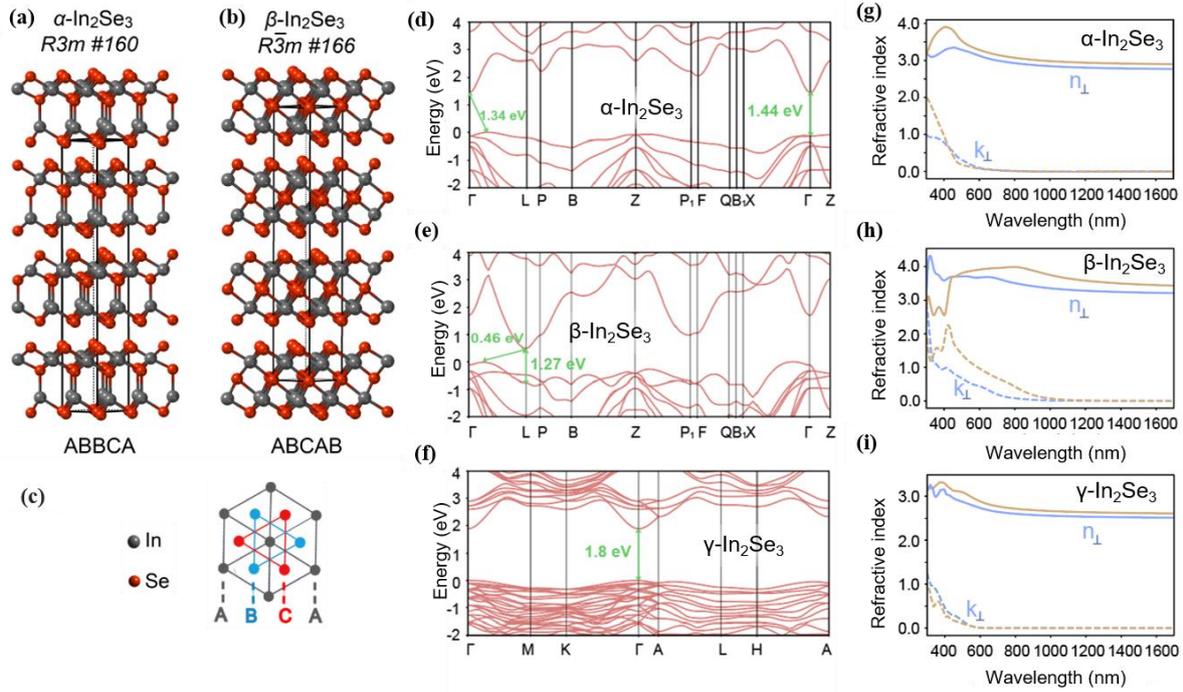

**Figure S1.** First principles DFT calculations of complex refractive index spectra. (a) Layered crystal structure of *α*- and (b) *β*-states In$_2$Se$_3$. (c) The quintuple layer stacking. (d) Electronic band structure of *α*- (e) *β*- and (f) *γ*-state In$_2$Se$_3$. (d) and (e) are based on the respective rhombohedral primitive cell, and (f) is based on the hexagonal primitive cell. (g) Calculated real (solid curves) and imaginary (dashed curves) parts of refractive index spectra in UV-vis-IR wavelength range for *α*- (layered and anisotropic), (h) *β*- (layered and anisotropic), and (i) *γ*- (bulk and isotropic) states.

**S2. Calculated device parameters in hybrid MRR for photo-thermal induced nonvolatile switching**

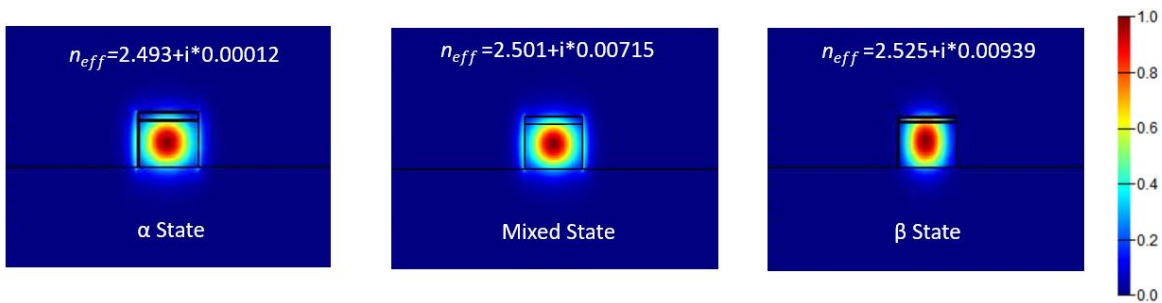

**Figure S2.** Optical mode profiles and correspondent effective index of the hybrid integrated waveguide at 1550 nm. The measured refractive index and corresponding film thickness are used for calculating the mode profile and effective index ($n_{eff}$). The thickness of the silicon waveguide is 220 nm with a width of 500 nm.

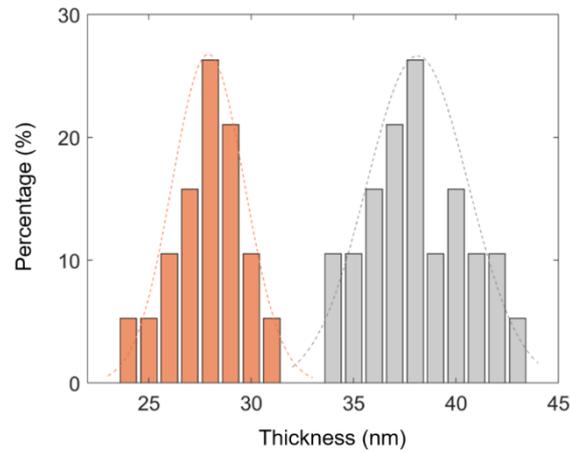

**Figure S3.** Statistical summary for the AFM measured thickness distribution in as prepared flakes (grey) and phase changed flakes by laser exposure (orange). The average thickness is 38.1 nm (including the transfer tape residual with average thickness of 5nm) for as prepared ($α$-state) flakes and 27.9 nm for phase changed flakes, with standard deviation ±3.0 nm and ±2.1 nm respectively. The average thickness transition of the MBE flakes is ~15±6%.

The numerical simulations for the photothermal effect are carried out in Lumerical DEVICE. The optical absorption calculated in DGTD solve utilizing the refractive index of $α$- and $β$-$In_2Se_3$ in Figure 3a&b. The incident pulses are 15 ns duration, with the exposure energy of 0.25 nJ and 0.56 nJ for $α$- and $β$-$In_2Se_3$, respectively. Then $In_2Se_3$ flake was set as the heat source in HEAT solver. Optical power was scaled according to the setup in our experiment. The thermal-related parameters are shown in Table S1. The thermal relaxation time constants are 5 ns and 17 ns for $α$- and $β$-states initiated phase transitions.

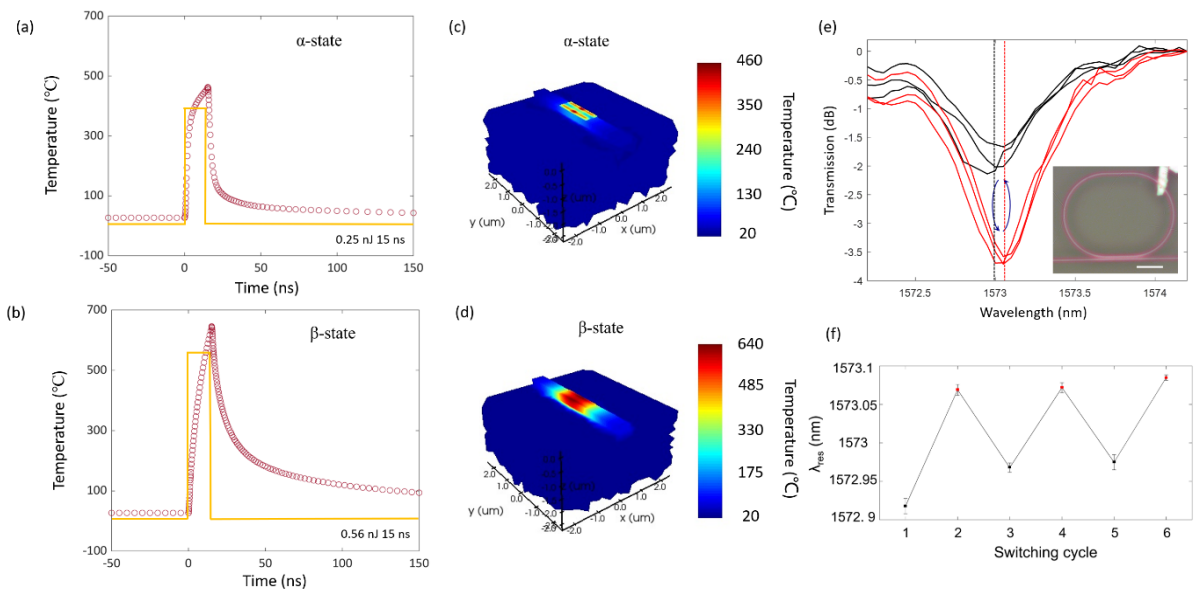

S-3

**Figure S4.** Simulated photo-thermal dynamics and temperature profiles of the $In_2Se_3$ on the waveguide. (a) Single-pulse excitation resulted in temperature variation for $α$- and (b) $β$-$In_2Se_3$, for normal incident optical pulse. (c) Three-dimensional temperature profile for $α$-$In_2Se_3$ and (d) $β$-$In_2Se_3$. (e) Measured MRR transmission spectra switched by 30 pulses. Black curves: $α$-$In_2Se_3$. Red curves: switched $β$-$In_2Se_3$. The average resonance shift between the two states (highlighted as verticle lines) is 120 pm. Inset: Optical micrscope image of the other device under test. The $In_2Se_3$ covered waveguide length is 1.5 μm. Scale bar: 5 μm. (f) Extracted resonance shift in Figure S4e.

**Table S1**. Thermal parameters used for the simulations.

| Parameters | $SiO_2$ | Si | $α$-$In_2Se_3$ | $β$-$In_2Se_3$ |
|---|---|---|---|---|
| Density(kg/m$^3$) | 2203 | 2330 | 5320 [a] | 6070 [a] |
| Heat capacity (J/kg/K) | 709 | 711 | 260 [S5] | 274 [S5] |
| Thermal conductivity(W/m/K) | 1.38 | 148 | 1 [S6] | 25 [S7] |

[a] Data source: https://materialsproject.org

We repeat the reversible nonvolatile switching in the other device (Figure S4e), where the $In_2Se_3$ covered length along the waveguide (Inset of Figure S4e) is close to the device in Figure 1 (1.5 μm). Transition into $β$-state was achieved at the exposure energy of 0.40 nJ. The measured resonance shift (>100 pm) and extinction ratio change (~2dB) are similar between the two devices. The phase change induced phase shift is calculated to be 0.038π, which derives the resonance shift of 148 pm, compared to measured value of 120 pm. When the exposure energy increased to around 0.65 nJ, the film changes back to $α$-state, the resonance peak shifted back, and the extinction ratio decreased to the original value. The retrieved transmission spectra in one state (marked in as the same color in Figure S4e) are not perfect overlapping, possibly due to reduced accumulation time for noise cancelation.

**S3. Derive linear and two-photon absorption coefficients from nonlinear transmission spectra in MRR**

To differentiate the contribution from the scattering loss and linear absorption loss, we utilize the integrated nonlinear spectroscopy to extract their loss rate by fitting with coupled-mode-theory [S8] (Figure S5). Here we compare the nonlinear transmission spectra of two devices with the same layout design in the silicon layer.

The nonlinear resonator transmissions with time-domain nonlinear coupled-mode-theory for the temporal rate evolution of the photon, carrier density, and temperatures as described by [S9-S11].

$$\frac{da}{dt} = (i(\omega_L - \omega_0 + \Delta\omega) - \frac{1}{2\tau_t})a + \kappa\sqrt{P_{in}} \tag{S-2a}$$

$$\frac{dN}{dt} = \frac{1}{2\hbar\omega_0\tau_{TPA}} \frac{V_{TPA}}{V_{FCA}^2} |a|^4 - \frac{N}{\tau_{FCA}} \tag{S-2b}$$

$$\frac{d\Delta T}{dt} = \frac{R_{th}}{\tau_{th}}(\frac{1}{\tau_{FCA}} + \frac{1}{\tau_{lin}})|a|^2 + \frac{\Delta T}{\tau_{th}} \tag{S-2c}$$

where $a$ is the amplitude of resonance mode; $N$ is the free-carrier density; $\Delta T$ is the cavity temperature shift. $P_{in}$ is the power carried by incident continuous-wave laser. $\kappa$ is the coupling coefficient between the waveguide and cavity. $\omega_L$-$\omega_0$ is the detuning between the laser frequency ($\omega_L$) and cold cavity resonance ($\omega_0$). The total cavity resonance shift is $\Delta\omega=\Delta\omega_N-\Delta\omega_T$, where $\Delta\omega_T=\omega_0\Delta T(dn/dT)/n$ is the thermal dispersion and $\Delta\omega_N=\omega_0\zeta N/n$ is the free-carrier dispersion, where $\zeta$ is the free-carrier dispersion coefficient. The total loss rate is $1/\tau_t = 1/\tau_{in}+1/\tau_v+1/\tau_{lin}+1/\tau_{TPA}+1/\tau_{FCA}$. $1/\tau_{in}$ and $1/\tau_v$ is the loss rates into the waveguide and vertical radiation into the continuum ($1/\tau_{in/v} =\omega/Q_{in/v}$).

The linear loss rate $1/\tau_{lin}$ represents the linear material absorption rate by the mid-gap defect states. The photo-thermal effect from those linear absorptions also contributes to photo-thermal dispersion (equation S-2c). The free-carrier absorption rate $1/\tau_{FCA}=c\sigma N(t)/n$. The field averaged two-photon absorption rate $1/\tau_{TPA}= \overline{\beta_2} \, c^2/n^2/V_{TPA}|a|^2$. The effective two-photon absorption coefficient ($\overline{\beta_2}$), mode volume ($V_{TPA}$), and effective mode volume for free carriers ($V_{FCA}$) can be calculated from the mode profiles (Figure S2). [S10] With a fixed set of parameters, the model fits the measured transmissions at increasing power levels for the microring resonator without (Figure S5a) and with $In_2Se_3$ layer cladding (Figure S5b). The key fitting parameters are listed in Table S2. The rest parameters used in equations S-2 are the same as in ref. S10.

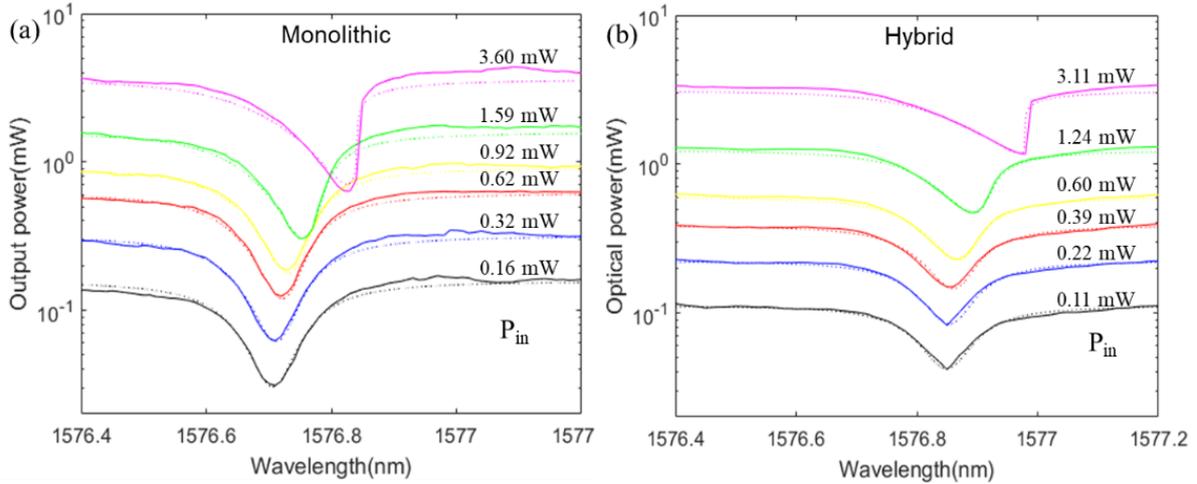

**Figure S5.** Nonlinear transmission spectra of (a) monolithic silicon MRR and (b) hybrid device with the same silicon layer geometry but 5 µm long In$_2$Se$_3$ coverage on the perimeter of the MRR. The solid curves are measurements, and the dashed curves are coupled-mode theory fittings.

**Table S2.** CMT fitting parameters for the nonlinear spectra of monolithic and hybrid MRRs with the same layout design and fabrication procedures

| Parameter | Symbol (unit) | Before transfer | After transfer |
|---|---|---|---|
| Effective TPA coefficient | $\beta_2$(cm/GW) | 1.5 | 6 [a] |
| **Linear absorption rate** | $1/\tau_{lin}$(GHz) | **16.7** | **20** [a] |
| Effective carrier lifetime | $\tau_{FCA}$(ns) | 0.5 | 0.7 [a] |
| Coupling quality factor | $Q_{in}$ | 17k | 28k |
| Intrinsic quality factor | $Q_v$ | 33k | 28k |
| **Derived scattering loss rate** | $1/\tau_v - 1/\tau_{lin}$(GHz) | 19.5 | 22.7 |

[a] Considering that only partial MRR is covered by PCM, the parameters are accumulation and average value over the parameter of the ring.

The intrinsic quality factors ($Q_v$) include contributions from both scattering and linear absorption, while only the linear absorption loss contributes to photo-thermal effect (Equation S-2c). In silicon, linear absorption is originated from the surface states. Since In$_2$Se$_3$ are layered materials, the surface defect density should be negligibly small. The extra linear material loss in the hyrid sample maybe originated from the Bi doping, which can be reduced by eliminating Bi diffusion in the MBE growth. The scattering loss is slightly higher than the linear material absorption (last row in Table S2). Compared to the monlithic device, the additional scattering loss from the In$_2$Se$_3$ flake is attributed to the organic residule from the thermal release tape (also discussed in figure S3).

Also, we extract the linear propagation loss determined $Q_v$ from the MRR before and after transferring, and summarize the flake covering length related $Q_v$ in table S3.

**Table S3.** Derived intrinsic quality factors of MRRs before and after transfer

| Device # | Coverage length | $Q_v$ (Before transfer) | $Q_v$ (After transfer) |
|---|---|---|---|
| 1 | 5 μm | 44,000 | 18,000 |
| 2 | 2 μm | 45,000 | 30,000 |
| 3 (Figure S8) | 2 μm | 33,000 | 28,000 |
| 4 (Figure 1) | 1.5 μm | NA | 4,800 |

**S4. Differential scanning calorimetry and In-situ high energy x-ray diffraction spectra results**

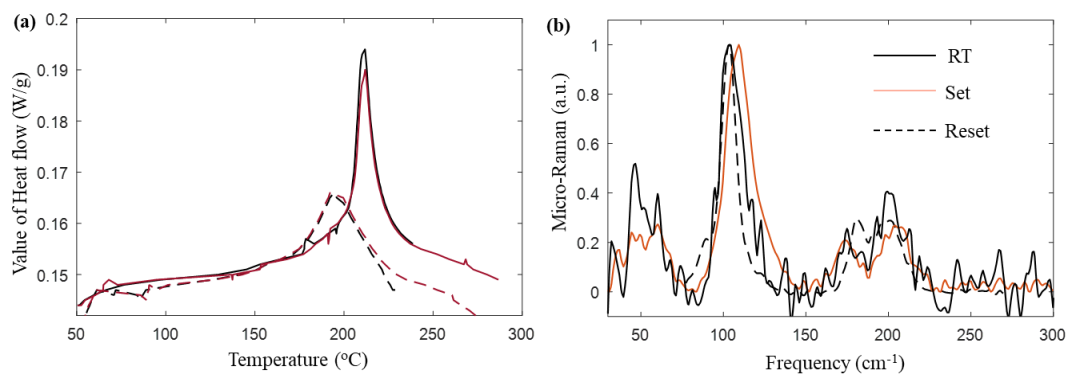

**Figure S6.** Cycles of differential scanning calorimetry measurement. (a) Absolution value of the heat flow during heating (solid curves) and cooling (dashed curves) at the first (black) and second (red) circles. (b) Correspondent micro-Raman spectra. The initial state sample is measured at room temperature (solid black curve). After DSC heating to 350°C and room temperature cooling, the shift in the micro-Raman spectrum indicates the transformation to β-state (orange curve). After a full cycle of heating and cooling in DSC, the sample resumes to α-state (dashed black curve). The characteristic DSC curve can be repeated for more than 10 cycles.

DSC measurement is used to characterize the thermal dynamics during phase transitions. DSC was conducted in repeated heating and cooling cycles at the ramp rate of 20°C (Figure S6a). Correspondent micro-Raman spectroscopies are taken for the powder before DSC, (black solid line in Figure S6b), after one heating process (red solid line in Figure S6b), and after one heating and cooling cycle (red dashed line in Figure S6b). The results show the overlapping DSC cycles in $In_2Se_3$.

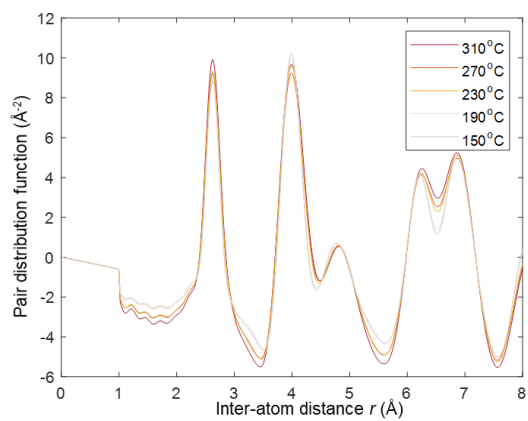

**Figure S7.** The temperature-dependent pair distribution functions (PDFs) with 40°C per step during cooling.

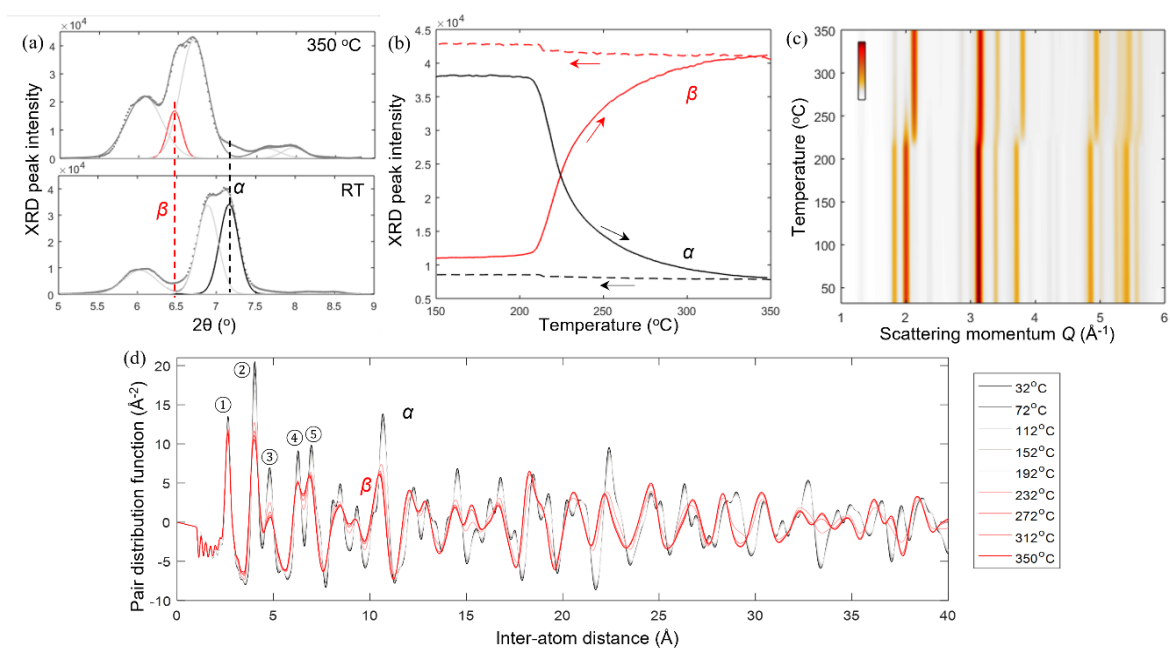

**Figure S8.** In-situ high energy X-ray diffraction. (a) Raw data of peak intensity versus angle. The peak marked in black disappears at the higher temperature, which features crystalline structure in *α*-state, with the rise of the red peak in *β*-state. (b) The peak intensity for *α*-state (black) and *β*-state (red) during heating (solid curve) and cooling (dashed curve). Those peaks are correspondent to the ones in panel (a). (c) Reduced structure factors *F*(*Q*) format versus scattering momentum and temperature. Significant spectra shift discontinuities are observed near 220°C, aligned with DSC measurement. (d) Converted PDF in the range from 0-40 Å. The temperature increases from 32°C to 350°C, with 40°C per step. The detailed analysis and comparison for the five primary peaks are given in Figure S9.

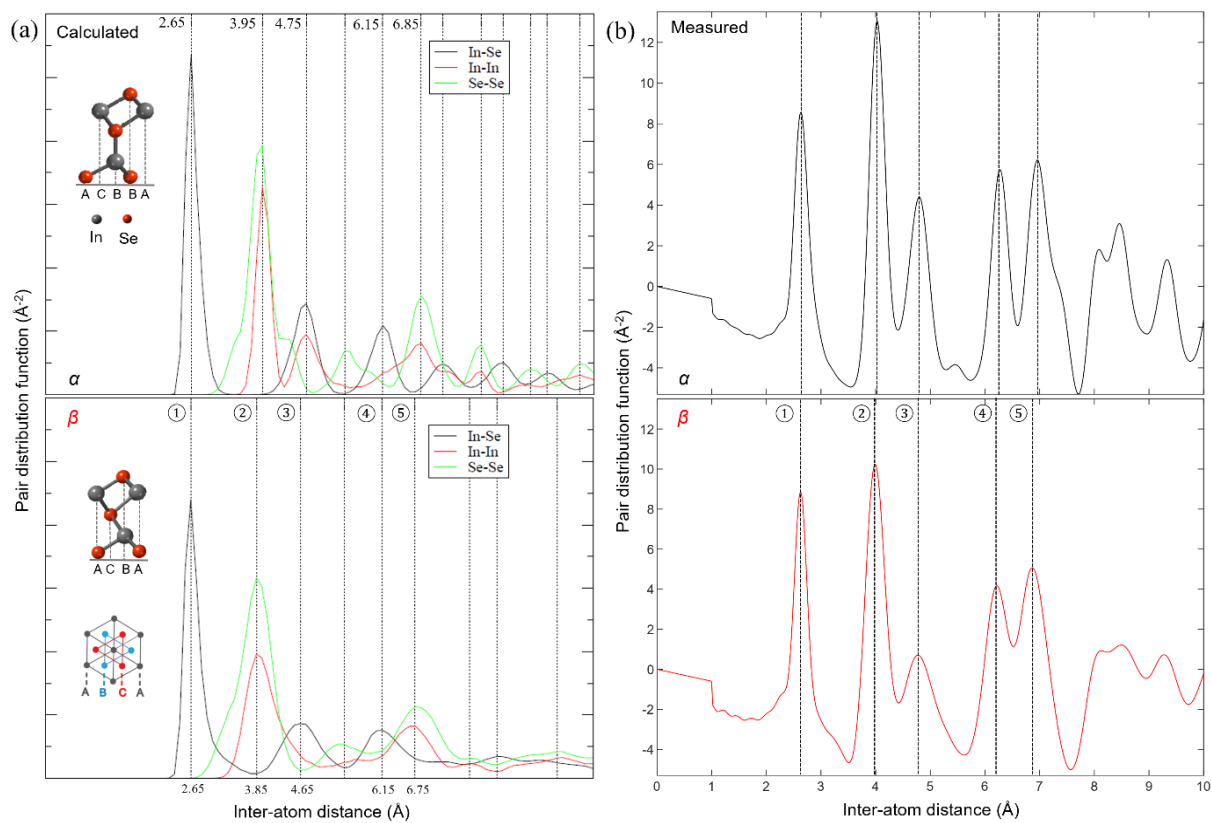

**Figure S9.** Comparison between calculated and measured PDFs. (a) Calculated distribution of inter-atom distances and (b) measured short-range PDFs for α- and converted β-In$_2$Se$_3$ at room temperature. One-to-one correspondence of the five primary peaks are identified in the figure, and both calculation and measurement share the same trend of bond lenth shift. Top/middle insets in a: correspondent atomic structure for α- (Wurtzite-type)/β-In$_2$Se$_3$ (face-centred cubic type). Bottom inset: top view of the system along the vertical direction. The atoms can be arranged in three (A, B, C) triangular lattice.

**S5. Molecular beam epitaxy grown thin-film In$_2$Se$_3$**

Molecular beam epitaxy (MBE) provides deterministic control of crystalline structure and quality in a scalable way [S11]. Here In$_2$Se$_3$ films are grown on c-plane sapphire in a dedicated Veeco GenXplor MBE chamber. Sapphire substrate has a lattice constant of 0.47 nm, which is close to the lattice constant of layered In$_2$Se$_3$ (0.4 nm). The successful growth of layered In$_2$Se$_3$ on the c-plane sapphire substrate has been demonstrated with Bi$_2$Se$_3$ buffer layer (lattice constant of 0.414 nm). It is noted that the substrate temperature and evaporation rate of each element (indium bulk source and selenium cracker source) to get the layered In$_2$Se$_3$. The altered growth condition can lead to the growth of γ-state In$_2$Se$_3$.

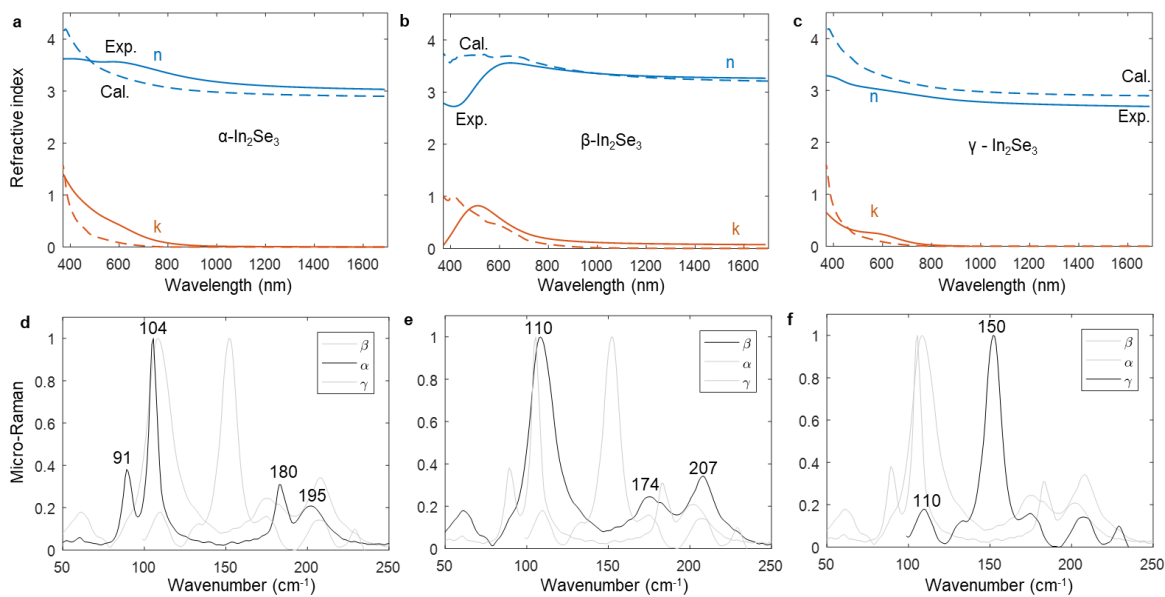

**Figure S10.** Measured complex refractive index spectra and correspondent Micro-Raman spectra in MBE grown $In_2Se_3$ thin film on a sapphire substrate (a) Real and imaginary refractive index of α-, (b) β-, and (c) γ-state $In_2Se_3$. The dashed curves are DFT calculations, and the solid curves are measurements. (d) (e) (f) Correspondent Micro-Raman spectra for sample in (d), (e) and (f), respectively.

A cracking source is used for selenium to improve selenium incorporation and reduce vacancies, while indium is evaporated using dual-filament effusion cells. The substrate temperature is monitored with a thermocouple. Sapphire substrates are first outgassed in the loadlock before being heated to 650°C in the MBE chamber to reduce contamination. The substrate temperature is then lowered for film growth. The selenium flux rate is held as a constant for all films, while the indium fluxes are adjusted using the effusion cell temperatures to achieve the right stoichiometry at a constant growth rate. It is difficult to nucleate a single polytype of $In_2Se_3$ directly from the sapphire substrate surface, therefore, we used a seed layer technique to grow films of a single polytype. We first grow 5 quituple layers (QL) of $Bi_2Se_3$ followed by 5 QL of $In_2Se_3$. Upon annealing at a high temperature, the two layers would potentially inter-diffuse and form a disdinguishable layer of $(Bi_{0.5}In_{0.5})_2Se_3$. We are then able to deposit lattice matched $In_2Se_3$ layer on this seed layer. The exact stoichiometry and spatial uniformity along the growth direction have not been fully characterized, but a clear interface between the seed layer and subsequent $In_2Se_3$ is observed in tunneling electron microscope image.

The crystal quality is monitored in situ by the reflection of high-energy electron diffraction (RHEED). All the films show similar streaky RHEED patterns, indicating single-phase and similar crystal quality. Given the low density of disordered states, those MBE grown crystalline

In$_2$Se$_3$ close to theoretical predictions. MBE growth requires a lattice match to the substrate. It is noted that the substrate temperature and evaporation rate of each element (indium bulk source and selenium cracker source) was carefully controlled to get the layered α-In$_2$Se$_3$. The altered growth condition can lead to the formation of γ-In$_2$Se$_3$, which will not be discussed here.

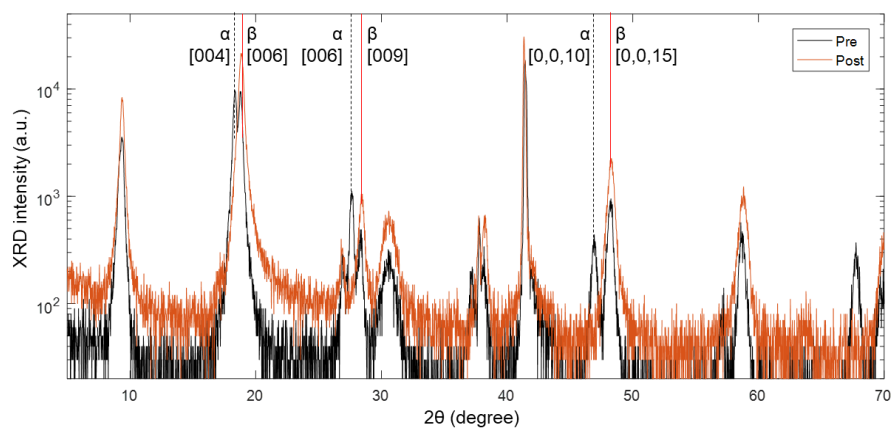

**Figure S11.** XRD spectra (5-70°) for as grown MBE In$_2$Se$_3$ sample on sapphire (black) and phase changed film (red) by annealing in an inert gas environment at 650°C for 5 min. Cu (Kα1) line is used as the XRD source, with a wavelength of 1.54 Å.

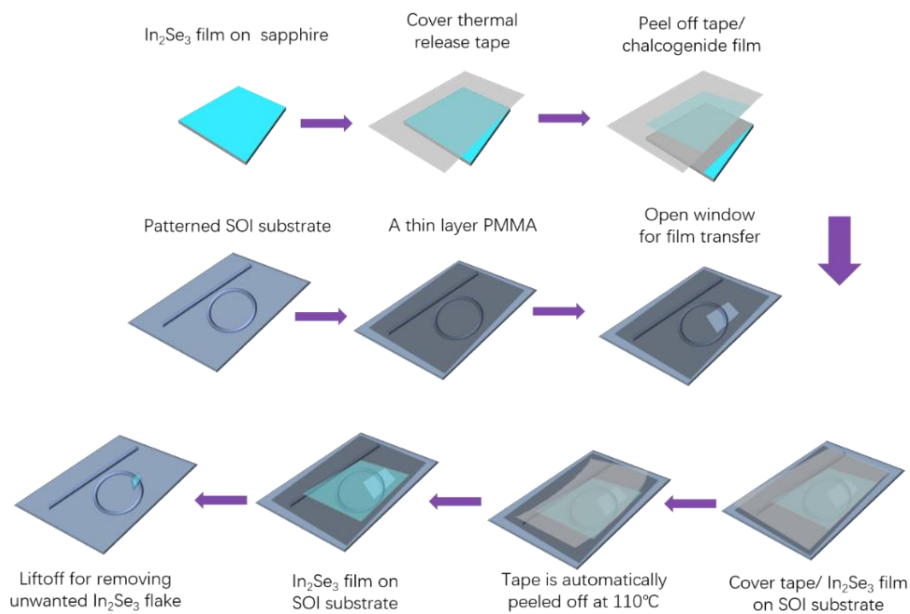

**Figure S12.** Schematic illustration of the exfoliation and transfer of MBE In$_2$Se$_3$ film to microring devices. The PMMA coating, lithography, development, and removal steps (for protecting waveguides and grating couplers) are not included here for clarity.

**S6. The transmission line method measured contact resistance**

Electronic devices are fabricated on as grown In$_2$Se$_3$ on the sapphire substrate to characterize dynamics of electronic properties during the phase transitions process. To differentiate the contributions from contact resistance/junction and the sheet resistance to the current-voltage curve, we used the transmission line method. Parallel metal electrodes with increasing gap width (from 0.5 to 8 µm) are fabricated on In$_2$Se$_3$ on a sapphire substrate (Figure S13a). The total resistance ($R_T$) versus gap distance ($L$) curve is shown in Figure S13b. The curve was fitted by TLM model [S12], the contact resistance with Cr-Au electrode is measured to be 7.87×10$^8$ Ω. The resistivity of the MBE-grown In$_2$Se$_3$ was 8.51×10$^5$ Ω·cm. This may be caused by the less impurity concentration of the MBE-grown scheme. Since the contact resistance with the Cr-Au electrode is measured to be an order of magnitude higher than channel resistance, most of the applied voltage drops on the contact region, which is not acceptable for the low power optoelectronic nonvolatile switch. Also, we observed metal migration into In$_2$Se$_3$ substrate with Au electrodes at high voltage bias. Wedge-shape electrodes were also patterned on the In$_2$Se$_3$-on-sapphire substrate for conducting the IV scan.

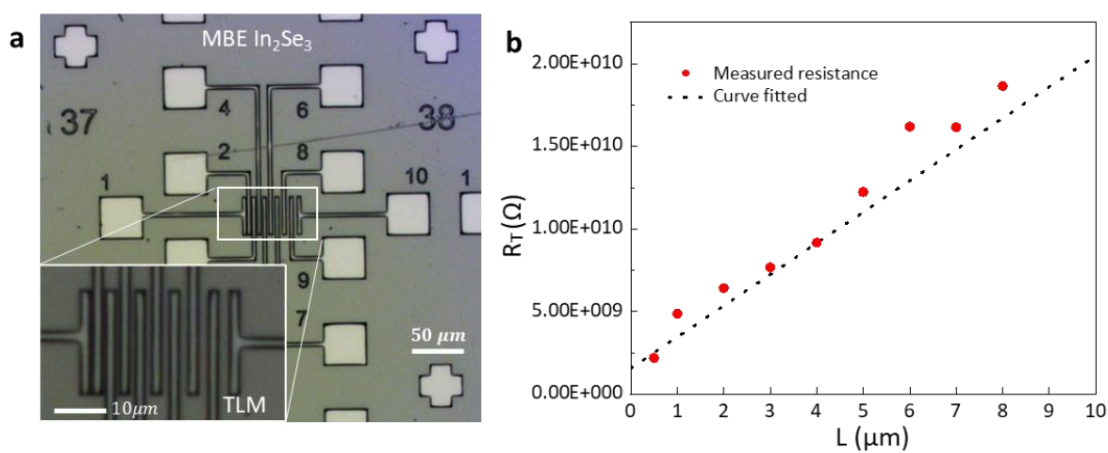

**Figure S13.** Transmission line method for extracting the contact resistance. (a) Optical image of fabricated electrodes for characterization sheet and contact resistance. (b) Resistance versus gap size plot extracting the sheet and contact resistance: $\rho_s$=8.51×10$^5$ Ω·cm, $R_c$=7.87×10$^8$ Ω respectively.

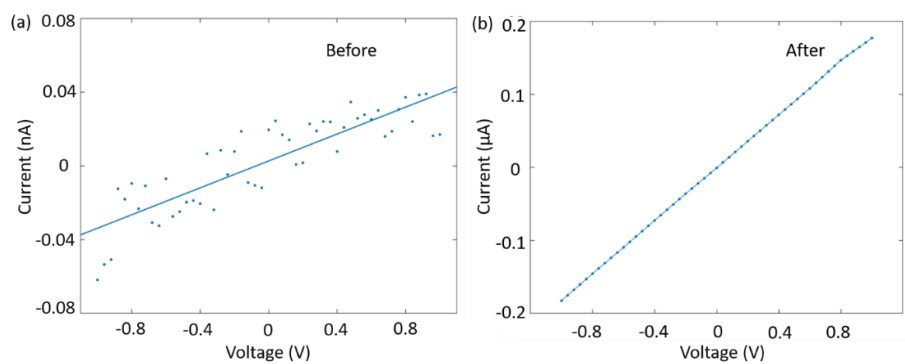

**Figure S14.** Low voltage I-V measurement before (a) and after (b) electrical switching as shown in the inset of Figure 4a. The fluctuation for the current in (a) was attributed to the limited accuracy of the source meter.

S-12

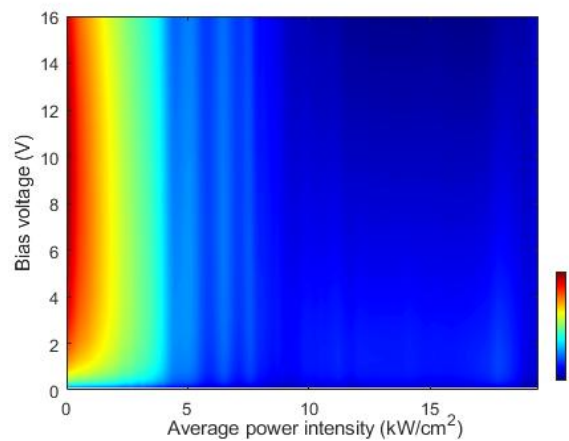

**Figure S15.** Source-drain bias and optical pulse intensity dependent resistance (log scale).

**S7. Reported phase transition temperatures in In$_2$Se$_3$**

As the phase transition temperature is highly sensitive to certain material impurities and stoichiometry. Those factors are sensitive to the material preparation procedures. Table S5 summarizes the reported phase transition temperatures and measured bandgaps.

**Table S4 | Measured phase transitions in In$_2$Se$_3$**

| Reference | Preparation | Phase transition temperatures | Bandgap |
|---|---|---|---|
| This work | Melt-quench | $\alpha$ - $\beta$ 207°C<br>$\beta$ - $\gamma$ 641.8°C<br>$\gamma$ - $\delta$ 692.8°C<br>melting: 885.6°C | $\alpha$: 1.44 eV (Optical)<br>$\beta$: 1.27 eV (Optical) |
| [S13] | Melt-quench | $\alpha$ - $\beta$ 200°C<br>$\beta$ - $\beta$' 200°C<br>$\beta$'- $\alpha$ 60°C | |
| [S14] | Melt-quench | $\alpha$ - $\beta$ 200°C<br>$\beta$ - $\gamma$ 520°C<br>$\gamma$ - $\delta$ 730°C<br>melting: 880°C | |
| [S15] | Melt-quench | $\alpha$ -$\gamma$ (VOSF) 570°C | $\alpha$: 1.26 eV<br>$\gamma$ (VOSF): 1.70 eV |
| [S16] | Cleaved from large single crystals | $\alpha$ - $\beta$ 200 °C | $\alpha$: 1.560 eV<br>$\beta$: 1.308 eV<br>$\gamma$: 1.812 eV |
| [S17] | Direct synthesis of the elements in an evacuated quartz tube | $\alpha$ - $\beta$ 200 °C<br>$\beta$ - $\gamma$ 650 °C | |
| [S18] | Exfoliate from Bulk | $\alpha$ - $\beta$' 300 °C | |
| [S19] | CVD grown | $\beta$ - $\beta$' -93 °C | $\beta$': 2.50 eV |
| [S20] | MBE grown on Au(111) | $\beta$'-$\beta$'' -196 °C | |
| [31] | Exfoliate from Bulk | $\alpha$ - $\beta$ 300 °C | $\beta$: 0.78 eV<br>$\gamma$: 1.86 eV |
| [62] | Melting the elements (from 600 °C to 800 °C) with rapid cooling (bismuth doped) | $\beta$ - another state 620 °C | |

WILEY-VCH## S8. Performance comparison between different optical phase-change materials

Table S5 compares the performance matrix reported for optical phase change materials. [4,5,11-14,15,17,S13,S18]: $Ge_2Sb_2Te_5$, $Ge_2Sb_2Se_4Te_1$, $Sb_2Se_3$ and $In_2Se_3$. The material parameters are usually characterized in the thin film of O-PCM, and the optical excitation induced phase transitions are measured in device levels. We listed all the information that can be found in the literature.

**Table S5.** Comparison of optical phase change materials and correspondent device performance

| Parameters | $Ge_2Sb_2Te_5$ | $Ge_2Sb_2Se_4Te_1$ | $Sb_2Se_3$ | $In_2Se_3$ |
|---|---|---|---|---|
| Material parameters | | | | |
| $|\Delta n|$ @1550 nm | 3.34 [13] | 2.0 [13] | 0.77 [5] | 0.45 (c) |
| k@1550 nm | $k_C$=1.882 [13] <br> $k_A$=0.192 [13] | $k_C$=0.350 [13] <br> $k_A$=1.8e-4 [13] | $k_C$ = ~0 [5] <br> $k_A$ = ~0 [5] | $k_\alpha$=0 (c) <br> $k_\beta$=0 (c) |
| Optical bandgap (eV) | ~0.7 [11] | 0.3 [4] | 1.6-2.1 [5] | $\alpha$ : 1.44 <br> $\beta$ : 1.27 |
| Phase change temperature (°C) | $T_{Melting}$ = ~620 [12] <br> $T_{crystallization}$ =200 [S21] | $T_{Melting}$>627 [13] <br> $T_{Crystallization}$~250 [13] /400 [4] | $T_{Melting}$ = 611 [14] <br> $T_{Crystallization}$ ~200 [5] | $T_{\alpha \to \beta}$ ~200 (m) <br> $T_{\beta \to \beta'}$~200 [S13] <br> $T_{\beta' \to \alpha}$~60 [S13] |
| Device parameters | | | | |
| Exposed laser energy or energy density | A→C: ~1 nJ [17] | C→A: 13.6 nJ [4] <br> A→C: 408 nJ [4] | C→A: ~5 nJ/μm² [5] <br> A→C: ~200 μJ/μm² [5] | $\alpha \to \beta$: ~5 nJ/μm² <br> $\beta \to \alpha$: ~11 nJ/μm² |
| Average power (mW) | A→C: >10 [17] | C→A: 136 [4] <br> A→C: 136 [4] | C→A: > 55 [5] <br> A→C: > 45 [5] | $\alpha \to \beta$: 4 <br> $\beta \to \alpha$: 8.7 |
| Spot size/ wavelength | Unknown spot size/ λ=658 nm | Unknown spot size/ λ=633 nm+780 nm | C→A: 1.5 μm, <br> A→C: 1.8 μm/ λ=638 nm | 10 μm / λ=1064 nm |
| Exposure time (ns) | C→A: < 1 [15] <br> A→C: $10^2$ [17] | C→A: $10^2$ [4] <br> A→C: $3 \times 10^6$ [4] a) | C→A: > $2 \times 10^2$ [5] <br> A→C: $10^7$ [5] | $\alpha \to \beta$: $10^5$ b) <br> $\beta \to \alpha$: $10^5$ b) |

a) A pulse train with a period of 1 μs, duty cycle of 0.03% (30 ns), and 100,000 repetitions was used.

b) A pulse train with a period of 3.3 μs, pulse duration of 15 ns, and 30 repetitions were used.

C: Crystalline; A: Amorphous; c: calculated; m: Measured.